\documentclass[12pt,reqno]{amsart}

\usepackage{a4}     
\usepackage{amsmath, amssymb}

\usepackage{graphicx}                      
\newcommand{\color}[2][{}]{}        

\usepackage{bbm}         

\theoremstyle{plain}            
\newtheorem{theorem}{Theorem}[section]
\newtheorem{proposition}[theorem]{Proposition}
\newtheorem{lemma}[theorem]{Lemma}
\newtheorem{corollary}[theorem]{Corollary}

\theoremstyle{definition}       

\theoremstyle{remark}
\newtheorem{remark}[theorem]{Remark}

\newcommand{\Sec}[1]{Section~\ref{sec:#1}}

\newcommand{\Thm}[1]{Theorem~\ref{thm:#1}}

\newcommand{\Lem}[1]{Lemma~\ref{lem:#1}}

\newcommand{\Lems}[2]{Lemmas~\ref{lem:#1} and~\ref{lem:#2}}

\newcommand{\Prp}[1]{Proposition~\ref{prp:#1}}

\newcommand{\Fig}[1]{Fig.~\ref{fig:#1}}

\numberwithin{equation}{section}

\DeclareMathOperator{\dom}    {dom}

\DeclareMathOperator{\supp}   {supp}

\DeclareMathOperator{\tr}     {tr}  

\newcommand{\spec}[2][{}]   {\sigma_{\mathrm{#1}}(#2)}

\newcommand{\ZZ}{\mathbb{Z}}
\newcommand{\RR}{\mathbb{R}}
\newcommand{\CC}{\mathbb{C}}

\newlength{\maxbreite}%
\newlength{\maxhoehe}%
\newlength{\maxtiefe}%

\newcommand{\stelldrueber}[3][0pt]{
  \settowidth{\maxbreite}{#3}%
  \settoheight{\maxhoehe}{#3}%
  \settodepth{\maxtiefe}{#2}%
  \addtolength{\maxhoehe}{\maxtiefe}%
  {\makebox[\maxbreite]{\raisebox{\maxhoehe}{\hspace{#1}#2}}%
  \makebox[0pt][r]{#3}}%
}

\newcommand{\overcirc}[1]       
{\stelldrueber[.45ex]{$\scriptscriptstyle\circ$}{${#1}$}}

\newcommand{\R}{\mathbb{R}} 

\renewcommand{\phi}{\varphi}   
\newcommand{\e}{\mathrm e}  
\newcommand{\im}{\mathrm i} 

\newcommand{\wt}{\widetilde}           

\newcommand{\Lsymb}    {L}     

\newcommand{\Lsqr}[2][{}]{\Lsymb_2^{#1}({#2})} 
\newcommand{\Lsqrloc}[2][{}]{\Lsymb_{2,\mathrm{loc}}^{#1}({#2})}

\newcommand{\normsqr}[2][{}]{\|{#2}\|^2_{#1}} 

\newcommand{\set}[2]{\{ \, #1 \, | \, #2 \, \} } 
\newcommand{\bigset}[2]{\bigl\{ \, #1 \, \bigl|\bigr. \, #2 \, \bigr\} }

\newcommand{\und}{\qquad\text{and}\qquad}

\newcommand{\Dir}{{\mathrm D}}              

\newcommand{\per}{\mathrm {per}}
\newcommand{\pp}{\mathrm {pp}}

\newcommand{\Ham}{H}  

\begin{document}
\title[Carbon nano-structures]{On the spectra of carbon nano-structures}

\author{Peter Kuchment}
\address{Mathematics Department,
Texas A\&M University College Station, TX 77843-3368 USA}
\email{kuchment@math.tamu.edu}

\author{Olaf Post}
\address{Institut f\"ur Mathematik,
         Humboldt-Universit\"at zu Berlin,
         Rudower Chaussee~25,
         12489 Berlin,
         Germany}
\email{post@math.hu-berlin.de}
\date{\today}




\begin{abstract}
An explicit derivation of dispersion relations and spectra for
periodic Schr\"{o}dinger operators on carbon nano-structures
(including graphene and all types of single-wall nano-tubes) is provided.
\end{abstract}

\maketitle

%
\section{Introduction}
\label{sec:intro}
%
Carbon nano-structures, in particular fullerenes (buckyballs),
carbon nano-tubes, and graphene have attracted a lot of attention
recently, due to their peculiar properties and existing or
expected applications (e.g., \cite{harris:02,Katsnelson,Saito}). Such
structures have in particular been modelled by quantum networks
(e.g., \cite{amovilli:04, alm:04b, korotyaev:06,
korotyaev:06_magn}), also called quantum graphs, which goes back
to quantum graph models in chemistry \cite{Pau,RuS} and physics
\cite{alexander:85, ARZ, deG, Montroll, Montroll2} (see also
\cite{BCFK,wrm,Kuch_thin} and references therein). A direct and 
inverse spectral study of Schr\"{o}dinger operators
on zig-zag carbon nano-tubes was conducted in \cite{korotyaev:06,
korotyaev:06_magn}.

In this paper, we take a different from \cite{korotyaev:06,
korotyaev:06_magn} approach to such a study. Namely, we provide a
simple explicit derivation of the dispersion relations for
Schr\"{o}dinger operators on the graphene structure, which in turn
implies the structure of the spectrum and density of states. This
derivation was triggered by the one done in \cite{KuKu99} for the
photonic crystal case, as well by \cite{amovilli:04,alm:04b},
albeit the presented computation is simpler and more convenient
for our purpose than the one in \cite{KuKu99}. It reflects the
known idea (e.g.,
\cite{alexander:85,ARZ,kuchment:04,kuchment:05,pankrashkin:pre05b})
that spectral analysis of quantum graph Hamiltonians (at least on
graphs with all edges of equal lengths) splits into two
essentially unrelated parts: analysis on a single edge, and then
spectral analysis on the combinatorial graph, the former being
independent on the graph structure, and the latter independent on
the potential.

The results are formulated in terms of the monodromy matrix (or rather
its trace, also called the Hill discriminant \cite{Eastham}) of the
$1D$ potential on one edge of the graphene lattice. Then this
dispersion relation, just by simple restriction procedure, gives
answers for any carbon nano-tube: zig-zag, armchair, or chiral. We
would like to emphasize that relations of properties of the
discriminant to the properties of the one-dimensional periodic
potential have been studied for a long time and are understood by now
extremely well (e.g., \cite{Mityagin, Eastham, garnett-trubowitz:84,
  garnett-trubowitz:87, IakStar, MagnusWinkl, March, Mckean,
  reed-simon-4, Novikov_solitons} and references therein). Thus, one
can extract all spectral information that might be needed, from our
explicit description of dispersion relations that involve the
discriminant.

The paper is structured as follows: in the next \Sec{structures}, the
main geometries and operators are introduced. \Sec{disp.rel} is
devoted to derivation of the dispersion relation and the spectral
structure for graphene, the main results provided in
\Thm{spec.graphen}. The following \Sec{spec.nano-tubes} deals with
nano-tubes. The main results accumulated in \Thm{spec.nano-tubes}
provide dispersion relations and all parts of the spectra of the
nano-tube operators. The last sections contain additional remarks and results, 
and acknowledgments.

%
\section{Schr\"{o}dinger operators on carbon nano-structures}
\label{sec:structures}
%
All structures studied in this text can be introduced through the
honeycomb graphene structure \cite{harris:02,Katsnelson,Saito}. So, we start
with discussing the latter.

\subsection{Graphen}
\label{ssec:graphen}

It is assumed that in graphene, the carbon atoms are situated at the
vertices of a hexagonal $2D$ structure $G$ shown in \Fig{hex-lattice}
below.
\begin{figure}[ht]
  \begin{picture}(0,0)%
    \includegraphics{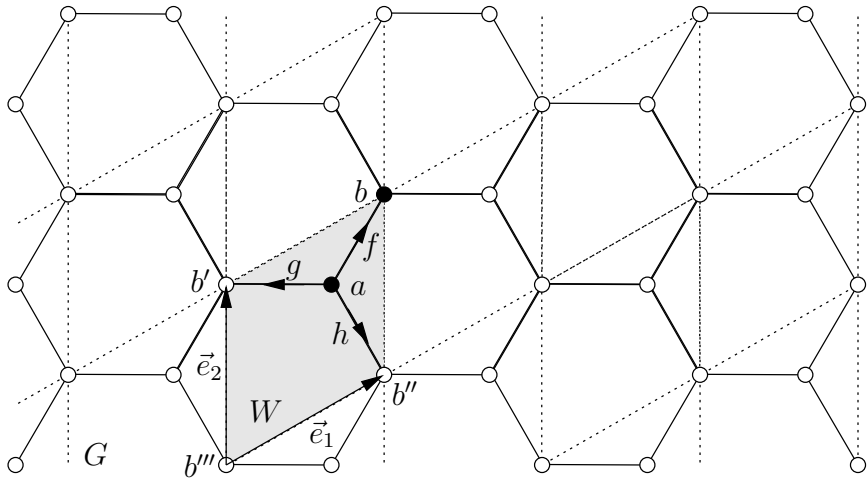}%
  \end{picture}%
  \setlength{\unitlength}{4144sp}%
  \begin{picture}(5146,2865)(83,-1973)
    \put(2386,-1501){$b''$}
    \put(2206,-601){$f$}
    \put(2161,-286){$b$}%
    \put(541,-1861){$G$}%
    \put(1182,-826){$b'$}%
    \put(2136,-844){$a$}%
    \put(1533,-1604){$W$}%
    \put(2026,-1136){$h$}%
    \put(1751,-717){$g$}%
    \put(1142,-1933){$b'''$}%
    \put(1216,-1321){$\vec e_2$}%
    \put(1891,-1726){$\vec e_1$}%
  \end{picture}%
  \caption{The hexagonal lattice $G$ and a fundamental domain $W$
    together with its set of vertices $V(W)=\{a,b\}$ and
    set of edges $E(W)=\{f,g,h\}$.}
  \label{fig:hex-lattice}
\end{figure}
We will assume that all edges of $G$ have length $1$. It will be
crucial for us to consider the following action of the group $\ZZ^2$
on $G$: it acts by the shifts by vectors
$p_1\vec{e}_1+p_2\vec{e}_2$, where $(p_1,p_2)\in\ZZ^2$ and vectors
$\vec e_1=(3/2,\sqrt 3/2)$, $\vec e_2=(0,\sqrt 3)$ are shown in
\Fig{hex-lattice}. We choose as a fundamental domain (Wigner-Seitz
cell) of this action the parallelogram region $W$ shown in the
picture. Here two vertices $V(W)=\{a,b\}$ are assumed to belong to
$W$, while the vertices $b'$, $b''$ and $b'''$ belong to other
shifted copies of the fundamental domain.  Three edges $f,g,h$
belong to $W$.  Although the graph $G$ does not have to be directed,
it will be convenient for us to assign directions to the edges in
$W$ as shown in the picture.

We will now equip the graph $G$ with the structure of a quantum graph
(quantum network) \cite{BCFK,KoS1,KotS,Kuch_thin,kuchment:04}.  This
requires introduction of a metric structure and of a differential
Hamiltonian. We assume that $G$ is naturally embedded into the
Euclidean plane, which induces the arc length metric on $G$, as well
as the identification of each edge $e$ in $G$ with the segment
$[0,1]$. Under this identification, the end points of an edge
correspond to the points $0$ and $1$ (this identification is unique up
to a symmetry with respect to the center of the edge, i.e., up to the
direction of the edge). We also obtain a measure (that we will call
$dx$), and the ability to differentiate functions along edges and to
integrate functions on $G$. In particular, the Hilbert space $\Lsqr G
:= \bigoplus_{e \in E(G)} \Lsqr e$ consisting of all square integrable
functions on $G$ can be naturally defined.  Here $E(G)$ denotes the
set of edges in $G$.

We now describe the graphene Hamiltonian $\Ham$ in $\Lsqr G$ that
will be studied in our paper (it has also been considered for some
special potentials in \cite{amovilli:04}). Such operators are used
for approximating the band structure of carbon nano-structures, as
well of other compounds (e.g., \cite{amovilli:04,alm:04b,RuS}, and
references therein). Let $q_0(x)$ be an $L_2$-function on the
segment $[0,1]$. We will assume that it is even:
\begin{equation}\label{eq:even}
q_0(x)=q_0(1-x).
\end{equation}
The evenness assumption is made not just for mathematical convenience. As 
the proposition below shows, this condition is required if one considers
operators invariant with respect to all symmetries of the graphene lattice.
One could consider a directed honeycomb graph, and thus 
avoid the evenness condition (hence losing invariance of the operator), but the 
authors did not see any compelling physical reason for doing so.
 
Using the fixed identification of the edges $e\in E(G)$ with $[0,1]$,
we can pullback the function $q_0(x)$ to a function (\emph{potential})
$q(x)$ on $G$. Notice, that due to the evenness condition imposed on
$q_0(x)$, the definition of the potential $q$ does not depend on the
orientations chosen along the edges. It is also easy to see that the
following claim holds:
\begin{proposition}
  \label{prop:symmetry}
  The potential $q$ defined as above, is invariant with respect to the
  full symmetry group of the honeycomb lattice $G$. Moreover, all
  invariant potentials from $\Lsqrloc G$ are obtainable by this
  procedure.
\end{proposition}
We skip the immediate proof of this statement.

We can now define our Hamiltonian $\Ham$. It acts along each edge $e$
as
\begin{equation}
  \label{eq:def.op}
  \Ham u(x) = -\frac{d^2u(x)}{dx^2} + q(x) u(x),
\end{equation}
where we use the shorthand notation $x$ for the coordinate $x_e$
along the edge $e$.

The domain $\dom \Ham$ of the operator $\Ham$ consists of the
functions $u$ that belong to the Sobolev space $H^2(e)$ on each
edge $e$ in $G$ and satisfy the inequality
\begin{equation}
  \label{eq:sobolev_finite}
    \sum\limits_{e\in E(G)}\normsqr[H^2(e)] u < \infty.
\end{equation}
They also must satisfy the so-called \emph{Neumann vertex conditions}
(also somewhat misleadingly called \emph{Kirchhoff vertex conditions})
at vertices. These conditions require continuity of the
functions at each vertex $v$ (and thus on all graph $G$) and vanishing of
the total flux, i.e.,
\begin{subequations}
  \label{eq:bd.cond}
  \begin{gather}
    u_{e_1}(v)=u_{e_2}(v) \qquad \text{if $e_1, e_2 \in E_v(G)$,}\\
    \sum_{e \in E_v(G)} u_e'(v) = 0
  \end{gather}
\end{subequations}
at each vertex $v \in V(G)$. Here $E_v(G):=\set{e \in E(G)}{v \in e}$
is the set of edges adjacent to the vertex $v$, $u_e$ is the
restriction of a function $u$ to the edge $e$, and $u_e'(v)$ denotes
the derivative of $u_e$ along $e$ in the direction away from the
vertex $v$ (\emph{outgoing direction}). Thus defined operator $\Ham$
is unbounded and self-adjoint in the Hilbert space $\Lsqr G$
\cite{KoS1,kuchment:04}. Due to the condition on the potential and Proposition 
\ref{prop:symmetry}, the
Hamiltonian $\Ham$ is invariant with respect to all symmetries of the
hexagonal lattice $G$, in particular with respect to the
$\ZZ^2$-shifts, which will play a crucial role in our considerations.

\subsection{Nano-tubes}
\label{sec:nano-tubes}
We provide here a very brief introduction to carbon nano-tubes. One can
find more detailed discussion and classification of nano-tubes in
\cite{harris:02, Saito}. We also emphasize that only single-wall
nano-tubes are considered.

Let $p\in \RR^2 \setminus \{\vec 0\}$ be a vector that belongs to the
lattice of translation symmetries of the honeycomb structure $G$. In
other words, $G+p=G$. Let us denote by $\iota_p$ the equivalence
relation that identifies vectors $z_1,z_2\in G$ such that $z_2-z_1$ is
an integer multiple of the vector $p$. A \emph{nano-tube} $T_p$ is the
graph obtained as the quotient of $G$ with respect to this equivalence
relation:
\begin{equation}
  \label{eq:nano-tube}
  T_p := G/\iota_p.
\end{equation}
This graph is naturally isometrically embedded into the cylinder
$\RR^2/\iota_p$.

\begin{figure}[h]
  \centering 
  \begin{picture}(0,0)%
    \includegraphics{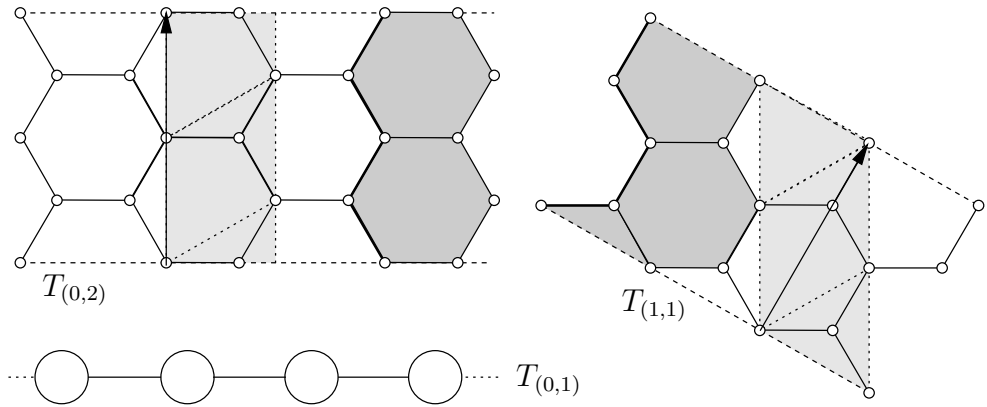}%
  \end{picture}%
  \setlength{\unitlength}{4144sp}%
  \begin{picture}(5877,2387)(-11,-1706)
    \put(3061,-1591){$T_{(0,1)}$}%
    \put(226,-1051){$T_{(0,2)}$}%
    \put(3691,-1141){$T_{(1,1)}$}%
\end{picture}%

  \caption{A \emph{zig-zag} (left) and \emph{armchair} (right)
    nano-tube $T_{(0,2)}$ and $T_{(1,1)}$, respectively. The vectors
    show the translation vector $p$. The name-giving fundamental
    domain of each of the nano-tubes is shaded in dark grey. The
    dashed lines have to be identified. Below, the (degenerate) zig-zag
    nano-tube $T_{(0,1)}$ is shown.}
  \label{fig:nano-tube}
\end{figure}

If $p=p_1\vec{e}_1+p_2\vec{e}_2$, we will abuse notations denoting
$T_p$ by $T_{(p_1,p_2)}$. For example, $T_{(0,N)}$ is the so-called
\emph{zig-zag nano-tube}, while $T_{(N,N)}$ is the so-called
\emph{armchair nano-tube}. The names come from the shape of the
boundary of a fundamental domain (cf.~\Fig{nano-tube}).
There are many other types of nano-tubes, besides the zig-zag and
armchair ones. They are usually called \emph{chiral}.

A degenerate example is given by the zig-zag nano-tube $T_{(0,1)}$,
which consists of a sequence of loops (``beads'') joined by edges into
a $1D$-periodic \emph{necklace structure} (see \Fig{nano-tube}).

One can notice that due to existence of rotational and mirror
symmetries of the hexagonal structure $G$, different vectors $p$ can
produce the same nano-tubes $T_p$. For instance,
$T_{(m,n)}=T_{(n,m)}$.  Also, zig-zag tubes $T_{(0,N)},T_{(N,0)}$, and
$T_{(N,-N)}$ are the same (see \cite{harris:02} and references therein
for a more detailed classification of nano-tubes).

The Hamiltonian $H_p$ on $T=T_p$ is defined exactly as for the
graphene lattice $G$. Alternatively, one can think of $\Ham$ acting on
functions on $G$ that are periodic with the period vector $p$ (this
definition requires some precision, since such functions do not belong
to $\Lsqr G$).
%
\section{Spectra of graphene operators}
\label{sec:disp.rel}
%
Here we calculate the dispersion relation and thus all parts of the
spectrum of the graphene Hamiltonian $\Ham$ (see also
\cite{amovilli:04,alm:04b} for related considerations). One can
notice that the density of states is determined by the dispersion
relation, and thus when the latter is known, the former can be
determined as well \cite{AM,reed-simon-4}.

We apply now the standard Floquet-Bloch theory (e.g., \cite{Eastham,
  kuchment:book, reed-simon-4}) with respect to the $\ZZ^2$-action
that we specified before. This theory also holds in the quantum graph
case, (see, e.g., \cite{Exner_superlat, EG, kuchment:91, kuchment:05,
  Oleinik} and references therein). This reduces the study of the
Hamiltonian $\Ham$ to the study of the family of Bloch Hamiltonians
$H^\theta$ acting in $\Lsqr W$ for the values of the
\emph{quasimomentum} $\theta$ in the Brillouin zone $[-\pi,\pi]^2$.
Here the Bloch Hamiltonian $H^\theta$ acts the same way $\Ham$ does,
but it is applied to a different space of functions. Each function
$u=\{u_e\}$ in the domain of $H^\theta$ must belong to the Sobolev
space $u_e \in H^2(e)$ on each edge $e$ and satisfy the vertex
conditions~\eqref{eq:bd.cond}, as well as the \emph{cyclic conditions}
(Floquet-Bloch conditions)
\begin{equation}
  \label{eq:floquet}
  u(x+p_1\vec{e}_1 + p_2\vec{e}_2) =
  \e^{\im p \cdot \theta} u(x) =
  \e^{\im (p_1 \theta_1 + p_2 \theta_2)} u(x)
\end{equation}
for any vector $p=(p_1,p_2) \in \ZZ^2$ and any $x\in G$.

Due to the conditions~\eqref{eq:floquet}, functions $u$ are uniquely
determined by their restrictions to the fundamental domain $W$. Then
conditions~\eqref{eq:bd.cond} and~\eqref{eq:floquet} reduce to
\begin{equation}
  \label{eq:bd.cond.th}
  \begin{cases}
    u_f (0)  = u_g (0)=u_h(0) =: A\\
     u_f' (0)+ u_g' (0)+ u_h'(0)=0\\
     u_f (1) =\e^{\im \theta_1} u_g(1)=\e^{\im \theta_2} u_h(1)
                                                              =:B\\
     u_f'(1) +\e^{\im \theta_1} u_g'(1)+ \e^{\im \theta_1} u_h'(1)=0.
   \end{cases}.
\end{equation}
By standard arguments (see e.g.~\cite[Theorem 18]{kuchment:04}),
$H^\theta$ has purely discrete spectrum
$\sigma(H^\theta)=\{\lambda_j(\theta)\}$. The graph of the multiple
valued function $\theta \mapsto \{\lambda_j(\theta)\}$ is known as
the \emph{dispersion relation}, or \emph{Bloch variety} of the
operator $\Ham$. It is known \cite{Eastham, IakStar, kuchment:book,
reed-simon-4} that the range of this function is the spectrum of
$\Ham$:
\begin{equation}
  \label{eq:range}
  \sigma (\Ham) = \bigcup_{\theta \in [-\pi,\pi]^2} \sigma(H^\theta).
\end{equation}
It is also well known that the dispersion relation determines not only
the spectrum, but the density of states of $\Ham$ as well
\cite{AM,reed-simon-4}.

So, our goal now is the determination of the spectrum of $H^\theta$
and thus the dispersion relation of $\Ham$. In order to determine this
spectrum, we have to solve the eigenvalue problem
\begin{equation}
  \label{eq:ew}
  H^\theta u = \lambda u
\end{equation}
for $\lambda\in \R$ and a non-trivial function $u \in \Lsqr W$
satisfying the above boundary conditions.

Let us denote by $\Sigma^\Dir$ the spectrum of the Dirichlet
Hamiltonian $H^\Dir$ acting as in \eqref{eq:def.op} on $(0,1)$ with
boundary conditions $u(0)=u(1)=0$.  If $\lambda \notin \Sigma^\Dir$,
there exist two linearly independent solutions $\phi_0$, $\phi_1$
(depending on $\lambda$) of the equation
\begin{equation}
  \label{eq:ode}
  -\phi'' + q_0 \phi = \lambda \phi
\end{equation}
on $(0,1)$, such that $\phi_0(0)=1$, $\phi_0(1)=0$ and $\phi_1(0)=0$,
$\phi_1(1)=1$. For example, if $q_0=0$ and $\lambda>0$ then we have
$\lambda \notin \Sigma^\Dir$ if and only if $\mu:=\sqrt \lambda \notin
\pi \ZZ$ and
\begin{equation}
  \label{eq:ansatz.ex}
  \phi_0(t) = \frac{\sin \mu(1-t)}{\sin \mu} \und
  \phi_1(t) = \frac{\sin \mu t}{\sin \mu}.
\end{equation}
We will often use the notation $\phi_{0,\lambda},\phi_{1,\lambda}$
to emphasize dependence on the spectral parameter.

We will assume that the functions $\phi_j$ are lifted to each of the
edges in $W$, using the described before identifications of these
edges with the segment $[0,1]$. Abusing notations, we will use the
same names $\phi_j$ for the lifted functions. Then, for $\lambda
\notin \Sigma^\Dir$ we can use~\eqref{eq:bd.cond.th} to represent any
solution $u$ of~\eqref{eq:ew} from the domain of $H^\theta$ on each
edge in $W$ as follows:
\begin{equation}
  \label{eq:ansatz.u}
  \begin{cases}
    u_f = A \phi_0 + B \phi_1  \\
    u_g = A \phi_0 + \e^{-\im \theta_1} B \phi_1\\
    u_h = A \phi_0 + \e^{-\im \theta_2} B \phi_1.
  \end{cases}
\end{equation}
With this choice, the continuity conditions in~\eqref{eq:bd.cond.th}
and the eigenvalue equation on each edge are automatically
satisfied. The remaining two equations that guarantee zero fluxes at
the vertices, lead to the system
\begin{equation}
  \label{eq:ew.system}
  \begin{cases}
    3\phi_0'(0) A +
         (1 + \e^{-\im \theta_1}+ \e^{-\im \theta_2} ) \phi_1'(0) B
                       = 0\\
    (1 + \e^{\im \theta_1} + \e^{\im \theta_2}) \phi_0'(1) A +
            3\phi_1'(1) B =0.
  \end{cases}
\end{equation}
Using the symmetry~\eqref{eq:even} of the potential $q_0$, we
obtain
\begin{equation}
  \label{eq:symm.der}
  \phi_1'(1) = - \phi_0'(0) \und
  \phi_0'(1) = - \phi_1'(0).
\end{equation}
Thus, the system~\eqref{eq:ew.system} reduces to
\begin{equation}
   \label{eq:ew.system2}
  \begin{cases}
   - 3\phi_1'(1) A +
         (1 + \e^{-\im \theta_1}+ \e^{-\im \theta_2} )  \phi_1'(0)B
                = 0\\
    (1 + \e^{\im \theta_1} + \e^{\im \theta_2}) \phi_1'(0) A -
           3 \phi_1'(1) B =0.
  \end{cases}
\end{equation}
The quotient
\begin{equation}
  \label{eq:def.eta}
  \eta(\lambda) :=
  \frac{\phi_{1}^\prime(1)} {\phi_{1}^\prime(0)} =
  \frac{\phi_{1,\lambda}^\prime(1)} {\phi_{1,\lambda}^\prime(0)}
\end{equation}
is well defined, since $\phi_{1}^\prime(0)\neq 0$. Thus, the
system~\eqref{eq:ew.system2} can be rewritten as
\begin{equation}
  \label{eq:ew.system3}
  \begin{cases}
   - 3\eta(\lambda) A +
         (1 + \e^{-\im \theta_1}+ \e^{-\im \theta_2} )  B
                = 0\\
    (1 + \e^{\im \theta_1} + \e^{\im \theta_2}) A -
           3 \eta(\lambda)B =0.
  \end{cases}
\end{equation}
The determinant of this system is clearly equal to
\begin{equation}
  \label{eq:determinant}
  \begin{array}{c}
    |1 + \e^{\im \theta_1}+ \e^{\im \theta_2}|^2-9\eta^2(\lambda) \\=
    3+2\cos{\theta_1}+2\cos{\theta_2}+2\cos{(\theta_1-\theta_2)}
      -9\eta^2(\lambda) \\=
    1+8\cos \frac{\theta_1-\theta_2}{2} \cos \frac{\theta_1}{2}
                                        \cos \frac{\theta_2}{2}
          -9\eta^2(\lambda).
  \end{array}
\end{equation}

Thus, we have proven the following lemma:
\begin{lemma}
  \label{lem:disp.rel}

  If $\lambda \notin \Sigma^\Dir$, then $\lambda$ is in the spectrum
  of the graphene Hamiltonian $\Ham$ if and only if there is $\theta \in
  [-\pi,\pi]^2$ such that
  \begin{equation*}
    9\eta^2(\lambda)=|1 + \e^{\im \theta_1}+ \e^{\im \theta_2}|^2,
  \end{equation*}
  or
  \begin{equation}
    \label{eq:dispersion}
    9\eta^2(\lambda) = 
    1+8\cos{\frac{\theta_1-\theta_2}{2}}
            \cos{\frac{\theta_1}{2}}\cos{\frac{\theta_2}{2}}.
  \end{equation}
\end{lemma}

\begin{remark}
  Note that this lemma gives a nice relation between the spectrum of
  the metric graph Hamiltonian $H$ and the discrete graph Laplacian
  \cite{Chung, Col} $\Delta U:= \frac 1 {\deg v} \sum_{w \sim v}
  U(w)$. Indeed, the Bloch Laplacian $\Delta^\theta$ on
  $\ell_2(\{a,b\}) \cong \CC^2$ has the matrix
        \begin{equation*}
                \Delta^\theta \cong \frac 13
                \begin{pmatrix}
                  0 & 1+\e^{-\im \theta_1} + \e^{-\im \theta_2}\\
                  1+\e^{\im \theta_1} + \e^{\im \theta_2} & 0
                \end{pmatrix}.
        \end{equation*}
  Thus, the lemma is just the statement that for $\lambda \notin
\Sigma^\Dir$, we have $\lambda \in \spec {H^\theta}$ if and only
if $\eta(\lambda) \in \sigma(\Delta^\theta)$, and hence $\lambda
\in \spec {H}$ if and only if $\eta(\lambda) \in \sigma(\Delta)$.

This relation between quantum and combinatorial graph operators is
well known and has been exploited many times (e.g.,
\cite{alexander:85, ARZ, Cat, kuchment:04, kuchment:05, KuchVain,
pankrashkin:pre05b}).
\end{remark}

\Lem{disp.rel}, in particular, says that in order to find the spectrum
of $\Ham$, we need to calculate the range of the following function on
$[-\pi,\pi]^2$:
\begin{equation}
  \label{eq:def.ew.func}
  F(\theta_1,\theta_2) =
     \sqrt{1+8\cos{\frac{\theta_1-\theta_2}{2}}
                \cos{\frac{\theta_1}{2}}\cos{\frac{\theta_2}{2}}}.
\end{equation}
\begin{lemma}
  \label{lem:range.g}
  The function $F$ has range $[0,3]$; its maximum is attained at
  $(0,0)$ and minimum at $(2\pi/3,-2\pi/3)$ and $(-2\pi/3,2\pi/3)$.
\end{lemma}
The proof of the lemma is straightforward, after noticing that the
function
\begin{equation*}
   F(\theta_1,\theta_2)^2=1 + 8\cos \frac{\theta_1-\theta_2}{2}
        \cos \frac{\theta_1}{2} \cos \frac{\theta_2}{2} =
  |1+\e^{\im\theta_1}+\e^{\im\theta_2}|^2
\end{equation*}
ranges from $0$ to $9$.

Next, we want to interpret the function $\eta(\lambda)$ in terms of
the original potential $q_0(x)$ on $[0,1]$. To this end, let us
extend $q_0(x)$ periodically to the whole real axis $\R$ and
consider the Hill operator $H^\per$ on $\R$ given as
in~\eqref{eq:def.op} with the resulting periodic potential:
\begin{equation}
  \label{eq:periodic}
  H^\per u(x) = -\frac{d^2u(x)}{dx^2}+q_0(x)u(x).
\end{equation}
(We maintain the notation $q_0(x)$ for the extended potential.)

We will be interested in the well studied spectral problem
\begin{equation}
   \label{eq:Hill}
   H^\per u=\lambda u
\end{equation}

As it is usually done in the theory of periodic Hill operators
(e.g.,~\cite{Eastham,IakStar, MagnusWinkl, Mityagin, reed-simon-4}),
we consider the so called \emph{monodromy matrix} $M(\lambda)$ of
$H^\per$. It is defined as follows:
\begin{equation*}
  \begin{pmatrix}
    \phi(1) \\ \phi'(1)
  \end{pmatrix}
  = M(\lambda)
  \begin{pmatrix}
    \phi(0) \\ \phi'(0)
  \end{pmatrix}
\end{equation*}
where $\phi$ is any solution of~\eqref{eq:Hill}. In other words, the
monodromy matrix shifts by the period along the solutions
of~\eqref{eq:Hill}. The matrix valued function $\lambda \mapsto
M(\lambda)$ is entire.

It is well known (see the references above) and easy to see that the
monodromy matrix has determinant equal to $1$. Its trace plays the
major role in the spectral theory of the Hill operator.  Namely, let
$D(\lambda)=\tr M(\lambda)$ be the so called \emph{discriminant} (or
Lyapunov function) of the Hill operator $H^\per$. The next
proposition collects well known results concerning the spectra of Hill
operators \cite{Eastham,IakStar, MagnusWinkl, Mityagin, reed-simon-4}:
\begin{proposition}
  \label{prp:hill}
  \indent
  \begin{enumerate}
  \item The spectrum $\sigma(H^\per)$ of $H^\per$ is purely absolutely
    continuous.

  \item $\sigma(H^\per)=\bigset{\lambda\in\R} {|D(\lambda)|\leq 2}$.

  \item $\spec{H^\per}$ consists of the union of closed
    non-overlapping and non-zero lengths finite intervals
    \emph{(bands)} $B_{2k}:=[a_{2k},b_{2k}]$,
    $B_{2k+1}:=[b_{2k+1},a_{2k+1}]$ such that
    \begin{equation*}
       a_0<b_0\leq b_1 <a_1\leq a_2 < b_2 \leq \ldots
    \end{equation*}
    and $\lim_{k \to \infty} a_k=\infty$.

    The (possibly empty) segments $(b_{2k},b_{2k+1})$ and
    $(a_{2k+1},a_{2k+2})$ are called the \emph{spectral gaps}.

    Here $\{a_k\}$ and $\{b_k\}$ are the spectra of the operator with
    periodic and anti-periodic conditions on $[0,1]$ correspondingly.

  \item Let $\lambda_k^\Dir\in\Sigma^\Dir$ be the $k$th Dirichlet
    eigenvalue, labelled in increasing order.  Then $\lambda_k^\Dir$
    belongs to (the closure of) the $k$th gap\footnote{If the gap
      closes, this boils down to a single point.}.  When $q_0$ is
    symmetric, as in our case, $\lambda_k^\Dir$ coincides with an edge
    of the $k$th gap\footnote{The same comment applies here.}.

  \item If $\lambda$ is inside the $k$th band $B_k$, then $D'(\lambda)
    \ne 0$, and $D(\lambda)$ is a homeomorphism of the band $B_k$ onto
    $[-2,2]$. Moreover, $D(\lambda)$ is decaying on $(-\infty,b_0)$
    and $(a_{2k},b_{2k})$ and is increasing on $(b_{2k+1},a_{2k+1})$.
    It has a simple extremum in each spectral gap $[a_k,a_{k+1}]$ and
    \ $[b_k,b_{k+1}]$.

  \item The dispersion relation for $H^\per$ is given by
    \begin{equation}
      \label{eq:disp_hill}
      D(\lambda)=2\cos \theta,
    \end{equation}
    where $\theta$ is the one-dimensional quasimomentum.
  \end{enumerate}
\end{proposition}
There are many other important direct and inverse spectral results
concerning the well studied (in particular, due to the inverse
scattering transform research) operator $H^\per$ (see,
e.g.,~\cite{Eastham, garnett-trubowitz:84, garnett-trubowitz:87,
IakStar, Mckean, MagnusWinkl, March, reed-simon-4,
Novikov_solitons}).

We will now see the relation between the function $\eta (\lambda)$
that was introduced for the graphene operator $\Ham$ and the
discriminant of $H^\per$. In order to do so, we introduce another
basis of solutions of~\eqref{eq:Hill}, namely $c_\lambda$ and
$s_\lambda$ with $c_\lambda(0)=1$, $c_\lambda'(0)=0$ and
$s_\lambda(0)=0$, $s_\lambda'(0)=1$ (the notations are chosen to
remind the cosine and sine functions in the case of zero potential).
Using this basis of the solution space, we obtain
\begin{equation}
  \label{eq:mon.matrix}
  M(\lambda) =
  \begin{pmatrix}
    c_\lambda (1) & s_\lambda (1)\\
    c_\lambda'(1) & s_\lambda'(1)
  \end{pmatrix}.
\end{equation}
A simple calculation (assuming the symmetry~\eqref{eq:even}) relates
the new basis with the one of $\phi_0$ and $\phi_1$
(which we now denote $\phi_{0,\lambda},\phi_{1,\lambda}$ to
emphasize dependence on the spectral parameter):
\begin{equation*}
  c_\lambda = \phi_{0,\lambda} + \eta(\lambda) \phi_{1,\lambda} \und
  s_\lambda = \frac 1 {\phi_{1,\lambda}'(0)} \phi_{1,\lambda}.
\end{equation*}
In particular, $c_\lambda(1)=s_\lambda'(1)=\eta(\lambda)$.  Thus,
\begin{equation}
  \label{eq:eta_delta}
  \eta(\lambda)=\frac 12 D(\lambda).
\end{equation}
For example, if $q_0=0$, then $\eta(\lambda)=\cos(\sqrt \lambda)$.

So far, we have been avoiding points of the Dirichlet spectrum
$\Sigma^\Dir$ of a single edge (i.e, the Dirichlet spectrum of the
potential $q_0(x)$ on $[0,1]$). We will now deal with exactly these
points.
\begin{lemma}
  \label{lem:ew.graphen}
  Each point $\lambda\in\Sigma^\Dir$ is an eigenvalue of infinite
  multiplicity of the graphene Hamiltonian $\Ham$. The corresponding
  eigenspace is generated by simple loop states, i.e., by
  eigenfunctions which live on a single hexagon and vanish at the
  vertices (see \Fig{wave} below).
\end{lemma}
\begin{proof}
  We need to guarantee first that each $\lambda \in \Sigma^\Dir$ is an
  eigenvalue. Indeed, an eigenfunction is provided by a simple loop
  state of the type shown for zero potential in \Fig{wave} below.
  \begin{figure}[ht]
    \label{fig:wave}
    \begin{center}
      \begin{picture}(0,0)%
        \includegraphics{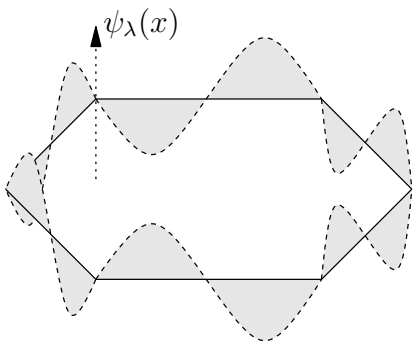}%
      \end{picture}%
      \setlength{\unitlength}{4144sp}%
      \begin{picture}(2454,1996)(799,-1520)
         \put(1396,344){$\psi_\lambda(x)$}%
       \end{picture}%
    \end{center}
    \caption{A simple loop state constructed from an odd eigenfunction
      on $[0,1]$.}
  \end{figure}
  It is constructed as follows. Let $\psi_\lambda(x)$ be an
  eigenfunction of the operator $-d^2/dx^2+q_0(x)$ with the eigenvalue
  $\lambda$ and Dirichlet boundary conditions on $[0,1]$.  Then, due
  to the symmetry (evenness) of the potential, we can assume the
  eigenfunction to be either even, or odd. For an odd eigenfunction
  $\psi_\lambda(x)$, repeating it on each of the six edges of a
  hexagon, we clearly get an eigenfunction of the operator $\Ham$. If
  it is an even eigenfunction, then repeating it around the hexagon
  with an alternating sign does the same trick.  Thus $\lambda \in
  \sigma_\pp(\Ham)$.

  It is well known then (e.g.,~\cite{Eastham}) that the multiplicity
  of the eigenvalue must be infinite. For completeness, we repeat here
  this simple argument. Let $M_\lambda\subset \Lsqr G$ be the
  eigenspace.  Consider a vector $\gamma$ that is a period of the
  lattice $G$ and the operator $S_\gamma$ of shift by $\gamma$ in
  $\Lsqr G$.  Then $S_\gamma$ acts in $M_\lambda$ as a unitary
  operator. If $\dim M_\lambda$ were finite, $S_\gamma$ would have had
  an eigenvector $f\in M_\lambda\subset \Lsqr G$ with an eigenvalue
  $\mu$, $|\mu|=1$.  However, such a function $f$ obviously cannot
  belong to $\Lsqr G$, since it is quasi-periodic in the direction of
  the vector $\gamma$.  This proves infinite multiplicity.

  We note now that due to~\cite{kuchment:91} (see
  also~\cite[Thm.~11]{kuchment:05}), linear combinations of compactly
  supported eigenfunctions are dense in the whole eigenspace
  $M_\lambda$. Thus, we only need to show that the simple loop states
  just described generate all compactly supported eigenfunctions in
  the space $M_\lambda$.

  Suppose that $\phi$ is a compactly supported eigenfunction of $\Ham$
  corresponding to the eigenvalue $\lambda\in\Sigma^\Dir$. First, note
  that $\phi$ vanishes at each vertex. Indeed, if this were not true,
  due to compactness of support, there would have been an edge such
  that at its one end $v_0$ (corresponding to $x=0$), $\phi(v_0)\ne
  0$, while at the other $v_1$ (corresponding to $x=1$), $\phi(v_1)=
  0$. Expanding into the basis $c_\lambda, s_\lambda$, we get
  \begin{equation*}
    \phi(x)=Ac_\lambda(x)+B s_\lambda(x).
  \end{equation*}
  In particular, $\phi(0)=A \ne 0$. Then $\phi(1)=A c_\lambda(1)=A
  s_\lambda'(1) \ne 0$, since $s_\lambda(1)=0$. This leads to a
  contradiction. Thus, $\phi$ vanishes at all vertices. In particular,
  on each edge it constitutes a Dirichlet eigenfunction for the Hill
  operator on this edge.

  This also implies that the support of $\phi$, as a graph, cannot
  have vertices of degree $1$ and $\supp \phi$ cannot be a tree. Thus,
  there must be a loop in the support of $\phi$. In particular, the
  outer boundary of the support must be a loop.  Take one boundary
  edge $e_0$. There is a hexagon inside the boundary loop which
  contains this edge. Consider a simple loop state $\phi_0$ coinciding
  with the eigenfunction $\phi$ on $e_0$ and extended to the hexagon
  the way it was described before. Subtracting $\phi_0$ from $\phi$,
  we obtain a new eigenfunction $\wt \phi$. The number of hexagons
  inside the boundary loop of the support of $\wt \phi$ is less than
  inside the support of $\phi$. Thus, continuing this procedure, we
  eventually represent $\phi$ as a combination of simple loop
  eigenstates.  \Fig{delete-proc} below illustrates this process.
  \begin{figure}[ht]
    \centering \includegraphics[scale=0.25]{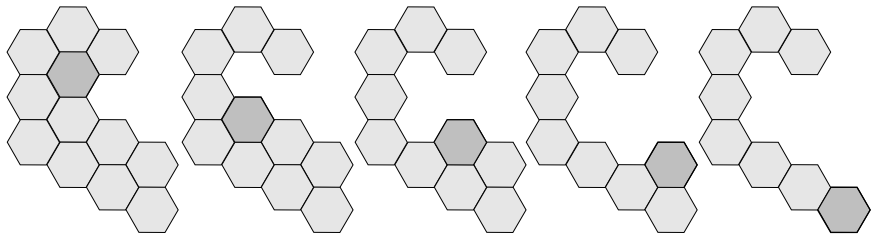}
    \caption{Deleting simple loop states (dark grey) from the support
      of an eigenfunction (light grey).}
      \label{fig:delete-proc}
  \end{figure}
\end{proof}

We can now completely describe the spectral structure of the
graphene operator $\Ham$.
\begin{theorem}
  \label{thm:spec.graphen}
  \indent
  \begin{enumerate}
  \item The singular continuous spectrum $\spec[sc] {\Ham}$ is empty.
  \item The absolutely continuous spectrum $\spec[ac] {\Ham}$ has
    band-gap structure and coincides as a set with the spectrum $\spec
    {H^\per}$ of the Hill operator $H^\per$ with potential $q_0$
    periodically extended from $[0,1]$.  In particular,
    \begin{equation*}
      \spec[ac]{\Ham}=\bigset{\lambda\in\R}{|D(\lambda)| \leq 2},
    \end{equation*}
    where
    $D(\lambda)$ is the discriminant of $H^\per$.

  \item The pure point spectrum $\spec[pp]{\Ham}$ coincides with
    $\Sigma^\Dir$ and thus, due to the evenness of the potential,
    belongs to the union of the edges of spectral gaps of
    $\spec{H^\per}=\spec[ac]{\Ham}$.

  \item The dispersion relation consists of the variety
    \begin{equation}\label{eq:dispersion_2}
        D(\lambda) = \pm \frac 23
        \sqrt{1 + 8\cos \frac{\theta_1-\theta_2}{2}
                   \cos \frac{\theta_1}{2} \cos \frac{\theta_2}{2}}
    \end{equation}
    and the collection of flat branches $\lambda\in\Sigma^\Dir$.

  \item Eigenvalues $\lambda\in\Sigma^\Dir$ of the pure point spectrum
    are of infinite multiplicity and the corresponding eigenspaces are
    generated by simple loop eigenstates.
  \end{enumerate}
  In particular, $\spec {\Ham}$ has gaps if and only if $\spec
  {H^\per}$ has gaps.
\end{theorem}
\begin{proof}
  \indent

  The claim~(i) about absence of the singular continuous spectrum is a
  general fact about periodic ``elliptic'' operators.  For instance,
  the standard proof applied for the case of periodic Schr\"odinger
  operators in \cite{reed-simon-4, Thomas} or \cite[Theorem
  4.5.9]{kuchment:book} works in our situation.  Alternatively,
  in~\cite{GerardNier} one can find this statement proven for a
  general case of \emph{analytically fibered operators}, which covers
  our situation as well.

  Statement~(iv) is a combination of \Lems{disp.rel}{ew.graphen} and
  formula~\eqref{eq:eta_delta}.

  The statement~(ii) about absolute continuity of the spectrum outside
  the points of $\Sigma^\Dir$ follows from~(iv) and the standard
  Thomas' analytic continuation
  argument~\cite{kuchment:book,reed-simon-4, Thomas}. We remind the
  reader that according to this argument, eigenvalues correspond to
  constant branches of the dispersion relation. It is clear that the
  dispersion curves~\eqref{eq:dispersion_2} have no constant branches
  outside $\Sigma^\Dir$.

  The claim~(iii) is just a combination of \Lem{ew.graphen} and of the
  Thomas' argument again, which shows that there are no eigenvalues
  outside $\Sigma^\Dir$.

  Finally,~(v) is a combination of~(iii) and \Lem{ew.graphen}.
\end{proof}

\begin{figure}[h]
  \centering
  \begin{picture}(0,0)%
    \includegraphics{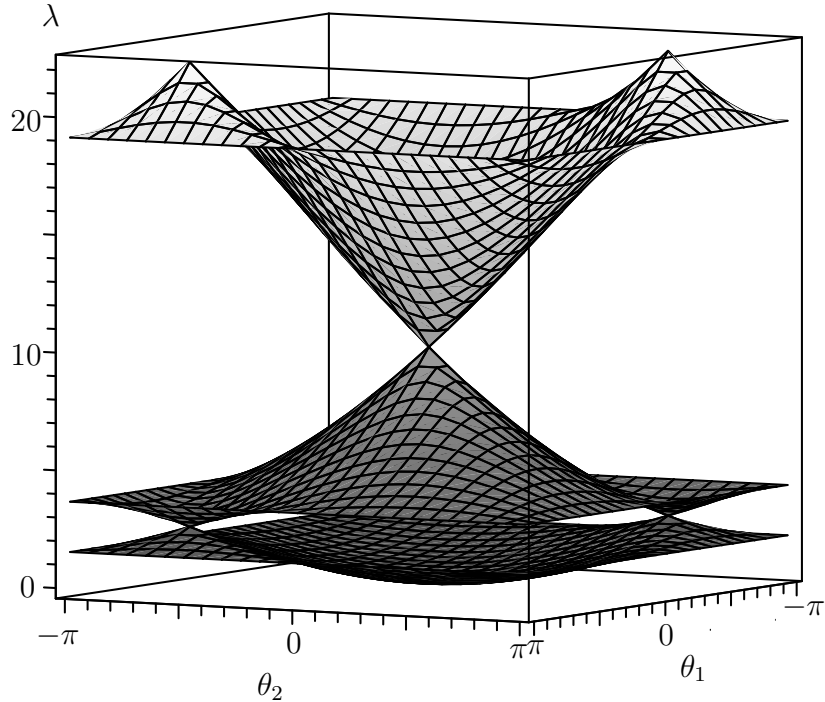}%
  \end{picture}%
  \setlength{\unitlength}{4144sp}%
  \begin{picture}(5110,5276)(856,-4932)
    \put(2528,-4153){$\theta_2$}%
    \put(5056,-4042){$\theta_1$}%
    \put(1107,-3568){$0$}%
    \put(1050,-2177){$10$}%
    \put(1050,-775){$20$}%
    \put(1206,-3848){$-\pi$}%
    \put(5666,-3675){$-\pi$}%
    \put(1240,-147){$\lambda$}%
    \put(4138,-3910){$\pi$}%
    \put(4036,-3931){$\pi$}%
    \put(4928,-3850){$0$}%
    \put(2700,-3900){$0$}%
  \end{picture}%
  \vspace*{-8ex} 

  \caption{The dispersion relation $\cos \sqrt \lambda = \pm
    F(\theta)/3$ for the free (i.e., $q_0=0$) case. The cones arising
    from $D(\lambda)=2\cos(\sqrt \lambda)=0$ are at the levels
    $\lambda=(\pi(2k+1))^2$. They are located \emph{inside} a band of
    the corresponding Hill operator. Note that the cones at
    $D(\lambda)=\pm 2$ (i.e., $\lambda=(\pi k)^2$ in the free case)
    are located at the \emph{band edges} of the Hill operator.}
  \label{fig:cones}
\end{figure}

It is clear that the function $F^2$ has non-degenerate minima $F=0$ at
the points $\theta=\pm (2\pi/3,-2\pi/3)$. Thus, the function $\pm F$
has conical singularities at these points. Since, according to
\Prp{hill}, the discriminant $D$ is monotonic with
non-degenerate derivative near each point where $D(\lambda)=0$, one
obtains the following conclusion:
\begin{corollary}
  The dispersion curve of the graphene operator $H$ has conical
  singularities at all spectral values $\lambda$ such that
  $D(\lambda)=0$.
\end{corollary}
These singularities (sometimes described as ``linear spectra'')
represent one of the most interesting features of graphene structures
(cf.~\Fig{cones}).
These singularities resemble Dirac spectra for
massless fermions and thus lead to unusual physical properties of
graphene (e.g., \cite{Katsnelson}). We see that quantum graph models
with arbitrary periodic potentials preserve this feature.

\section{Spectra of nano-tube operators}
\label{sec:spec.nano-tubes}

We use here the notations concerning nano-tubes that were introduced
in \Sec{nano-tubes}. Consider a vector $p=(p_1,p_2)\in\ZZ^2$ that
belongs to the lattice of translation symmetries of the graphene
structure $G$, i.e., it shifts the structure by
$p_1\vec{e}_1+p_2\vec{e}_2$ (see \Fig{hex-lattice}). We will use, as
before, the corresponding nano-tube $T_p=T_{(p_1,p_2)}$ and the
Hamiltonian $H_p=H_{(p_1,p_2)}$ on $T_p$ (see \Sec{nano-tubes}).

Let $B=[-\pi,\pi]^2$ be the Brillouin zone of graphene.  Then, as we
discussed in the previous section, the Floquet-Bloch theory provides
the direct integral expansion
\begin{equation}
  \label{eq:direct_int}
  \Ham = \int\limits_B^\oplus H^\theta d\theta.
\end{equation}
In the case of the nano-tube $T_p$, only the values of quasimomenta
$\theta$ enter that satisfy the condition
\begin{equation}
  \label{eq:line}
  p\cdot \theta = p_1 \theta_1 + p_2 \theta_2 \in 2\pi \ZZ
\end{equation}
since a function on $T_p$ lifts to a $p$-periodic function $u$ on
$G$, i.e.,
\begin{equation*}
  u(x+p_1 \vec e_1 + p_2 \vec e_2)=u(x)
\end{equation*}
(cf.~\eqref{eq:floquet}).  Thus, let us consider the following subset
$B_p\subset B$:
\begin{equation}
  \label{eq:briou}
  B_p := \bigset {\theta \in [-\pi,\pi]^2}
              {p\cdot \theta \in 2\pi \ZZ}.
\end{equation}
Then, we have the direct integral decomposition
\begin{equation}
  \label{eq:direct_int_tube}
  H_p = \int\limits_{B_p}^\oplus H^\theta d\theta.
\end{equation}
In particular, the spectrum of $H_p$ is given by
\begin{equation}
  \label{eq:spec.nano}
  \spec {H_p} = \bigcup_{\theta \in B_p} \spec {H^\theta},
\end{equation}
and the dispersion relation for $H_p$ is just the restriction to $B_p$
of the dispersion relation for $\Ham$ described in the part~(iv) of
\Thm{spec.graphen}.

\begin{figure}[ht]
 \begin{center}
    \begin{picture}(0,0)%
      \includegraphics{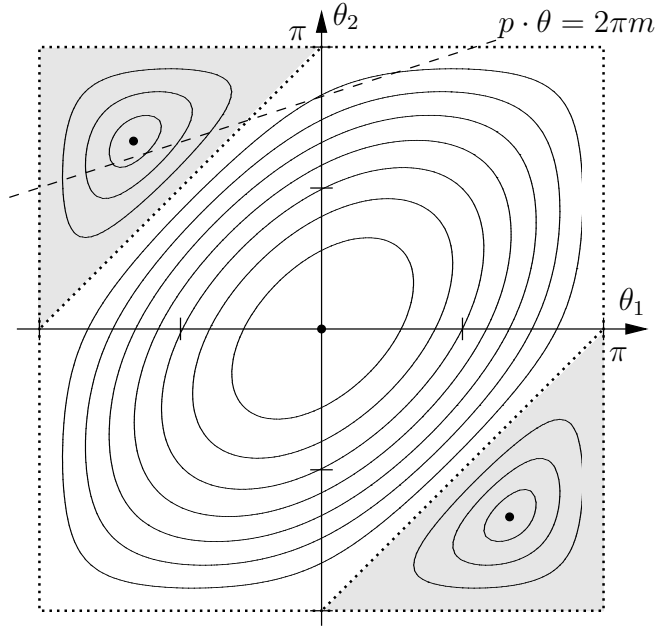}%
    \end{picture}%
    \setlength{\unitlength}{4144sp}%
    \begin{picture}(3849,3732)(34,-2863)
      \put(1981,749){$\theta_2$}
      \put(3691,-961){$\theta_1$}
      \put(1716,654){$\pi$}
      \put(3634,-1265){$\pi$}
      \put(2971,704){$p \cdot \theta=2\pi m$}
    \end{picture}

   \caption{The level curves of $F$ for levels varying from $0$ (the two dots at
     $\pm(2\pi/3,-2\pi/3)$) till $3$ (the dot at the origin).
     The
     level curve associated to $F=1$ is dotted, the areas where $F<1$
     are shaded. The dashed line is the line in $B_p$ closest to the
     minimum point $(-2\pi/3,2\pi/3)$ (in the case $p_1-p_2=3m \pm
     1$).}
  \label{fig:levelcurves}
  \end{center}
\end{figure}

This implies for instance that we still have $\Sigma^\Dir \subset
\spec[pp]{H_p}$ and the rest of the spectrum is determined by the
pre-image
\begin{equation}
  \label{eq:dispersion_p}
  \eta^{-1} \Bigl\{\pm \frac 13 F(B_p)\Bigr\} =
      D^{-1}\Bigl\{\pm \frac 23 F(B_p)\Bigr\}.
\end{equation}
One should notice that it is conceivable that non-constant branches of
the dispersion curves~\eqref{eq:dispersion} might sometimes have
constant restrictions to $B_p$, thus providing new eigenvalues for the
nano-tube Hamiltonian. This happens if the line~\eqref{eq:line} is a
level set of the function $F$. It is easy to find such lines.
\begin{lemma}
  \label{lem:level_lines}
  The only linear level sets of the function $F$ inside $B$ are the
  following ones:
  \begin{enumerate}
  \item $\theta_1=\pm \pi$

  \item $\theta_2=\pm \pi$

  \item $\theta_1-\theta_2=\pm \pi$
  \end{enumerate}
  On these lines $F(\theta_1,\theta_2)=1$.
\end{lemma}
The proof of the lemma is immediate from the
expression~\eqref{eq:def.ew.func} for the function $F$ (see also
\Fig{levelcurves} for the level sets of $F$, which illustrates this
statement).
We will see that existence of such lines leads to additional pure
point spectrum for some types of nano-tubes.

In order to determine the spectra of nano-tubes, we need to know the
ranges of the function $F$ (cf.~\eqref{eq:def.ew.func}) restricted to
$B_p$. These are described in the following

\begin{lemma}
  \label{lem:restrict}
    \indent
  \begin{enumerate}
  \item The function $F$ restricted to $B_p$, achieves its maximum $3$
    at $(0,0) \in B_p$ for any $p$.

  \item The location and value of the minimum depends on the vector
    $p$. The minimal value
    \begin{equation*}
    \label{eq:minimum}
      \alpha(p):=\mathop{\min}\limits_{\theta\in B_p}F(\theta)
    \end{equation*}
    satisfies
    \begin{equation}
    \label{eq:minimum_bound}
      \alpha(p)\in [0,1]
    \end{equation}
    for any $p$.

  \item $\alpha(p)=0$ if and only if $p_1-p_2$ is divisible by $3$.

  \item $\alpha(p)=1$ if and only if $p=(0,\pm 1)$, $(\pm 1,0)$,
    $(0,\pm 2)$, $(\pm 2,0)$, $(1,-1)$, $(-1,1)$, $(2,-2)$, or
    $(-2,2)$. (All these cases correspond to zig-zag nano-tubes).

  \item In all cases not covered by~(iii) and~(iv), let $p_1-p_2=3m\pm
    1$. Then $\alpha(p)\in (0,1)$ can be found by minimizing the
    function $F$ over the line $p_1\theta_1+p_2\theta_2=2\pi m$.

    In particular, in the case when $p=(0,N)$ with $N=3m\pm 1 >2$
    ($m\in \ZZ$), one has
    \begin{equation}
    \label{eq:alpha_0N}
      \alpha \bigl( (0,N) \bigr) =
      \Bigl| 2\cos{\frac{\pi m}{N}}-1 \Bigr|
    \end{equation}
    (this formula gives the correct answer $\alpha(p)=0$ also when
    $N=3m$).

  \item
     \begin{equation}
     \label{eq:alpha_limit}
      \lim\limits_{|p|\rightarrow\infty} \alpha(p)=0.
     \end{equation}
  \end{enumerate}
\end{lemma}
\begin{proof}
  \indent

  (i)~The claim about the maximum is straightforward, since the only
  maximum point $(0,0)$ of $F$ in $B$ belongs to $B_p$ for any $p$.

  (ii)~The expression~\eqref{eq:def.ew.func} shows that the set of
  points $\theta\in B$ where $F=1$, consists of four lines
  $\theta_j=\pm \pi$, as well as two lines $\theta_1-\theta_2=\pm
  \pi$. Since no line $p\cdot \theta=0$ can miss all these points, we
  conclude that $\alpha(p)\leq 1$ for any $p$. The inequality
  $\alpha(p)\geq 0$ is obvious, since, as we have already discovered,
  the expression under the square root in~\eqref{eq:def.ew.func} has
  its minimum equal to $0$.

  (iii)~As we have already indicated before, the points where $F$
  reaches its minimum are $(2\pi/3,-2\pi/3)$ and $(-2\pi/3,2\pi/3)$.
  Thus, for $\alpha(p)=0$ to hold, one of the lines $p\cdot\theta=2\pi
  n$ must pass through one of these points. Thus, $p_1-p_2=\pm 3n$,
  and the claim is proven.

  (iv) In order for $\alpha(p)$ to be equal to $1$, the lines
  $p\cdot\theta\in 2\pi\ZZ$ should not enter the zones where $F<1$
  (shaded in \Fig{levelcurves}). It is clear that when $p_1,p_2$ are of
  the same sign, this is impossible for the line $p\cdot\theta=0$,
  unless one of the coordinates $p_j$ is equal to zero. One can assume
  then, due to symmetries, that $p=(0,N)$, $N>0$. In this case, the
  line $p\cdot\theta=N\theta_2=2\pi$ enters the shaded region, unless
  $N\leq 2$.

  If the coordinates $p_j$ have opposite signs, then in order for the
  first ``non-trivial'' line $p\cdot\theta=\pm 2\pi$ not enter the
  Brillouin zone (and thus in particular the shaded area), one has to
  satisfy the condition $|p_1\pi-p_2\pi|\leq 2\pi$. Due to the signs
  of $p_j$s being opposite, this means that $|p_1|+|p_2|\leq 2$. This
  restricts the situation to the vectors $p=(1,-1)$ and $(-1,1)$ only,
  which do satisfy $\alpha(p)=1$. If this line does enter the
  Brillouin zone, the only case when the shaded area is not entered is
  when $p=(2,-2)$ or $(-2,2)$ and thus the line goes along the
  boundary of the shaded region.

  (v)~In order to find $\alpha(p)$, we need to minimize $F$ over the
  set of lines $p\cdot \theta=2\pi n$ for all such integers $n$ that
  the line intersects the Brillouin zone $B$. This entails first
  determining the appropriate of value of $n$ and then minimizing over
  the corresponding line. The minima of $F$ are located at the points
  $(-2\pi/3,2\pi/3)$ and $(2\pi/3,-2\pi/3)$ and are the only local
  minima in the shaded regions shown in \Fig{levelcurves}. Thus, we
  need to find the value of $n$ that provides a line closest to a
  minimum point (see again \Fig{levelcurves}). Evaluating
  $p\cdot\theta$ at the point $(-2\pi/3,2\pi/3)$, we get $2\pi
  (p_1-p_2)/3=2\pi (m\pm 1/3)$. This suggests that the line
  $p\cdot\theta=2\pi m$ is the right one.

  When $p=(0,N)$ with $N=3m\pm 1>2$, we conclude that we need to
  minimize $F$ over the line $N\theta_2=2\pi m$. Substituting the
  value $\theta_2=2\pi m/N$ into the modified expression for $F$,
  \begin{equation}
    \label{eq:F_modified}
     F(\theta_1,\theta_2) =
     \sqrt {1+4\cos \frac {\theta_2} 2
        \Bigl[(\cos \frac{\theta_2} 2  +
               \cos \Bigl(\theta_1 - \frac{\theta_2} 2 \Bigr)
        \Bigr]},
  \end{equation}
  leads to a simple minimization with respect to $\theta_1$ and thus
  to the formula~\eqref{eq:alpha_0N}.

  (vi)~This claim is clear, since when $|p|\rightarrow\infty$, the
  lines $p\cdot\theta\in2\pi\ZZ$ form a dense set in $B$, and thus the
  minimum of $F$ over $B_p$ approaches zero.
\end{proof}

Let us now concentrate for a moment on the additional pure point
spectrum that arises due to the linear level sets described in
\Lem{level_lines}. Let us assume that $p=(0,2N)$, $N\in\ZZ$.  Then,
according to that lemma, the line $p\cdot\theta=2N\pi$ is a level $1$
set of $F(\theta)$. Consider $\lambda$ such that $\eta(\lambda)=1/3$
(or $\eta(\lambda)=-1/3$). We will now construct a compactly supported
eigenfunction for $H_{(0,2N)}$. In order to do so, let us notice that
$\eta(\lambda)=1/3$ means that $\phi^\prime_{1,\lambda}(0) =
3\phi^\prime_{1,\lambda}(1)$. Let us construct a function $\phi_{+}$
on the boldface structure $Z$ in \Fig{extra-ef} below.
\begin{figure}[ht]
    \begin{picture}(0,0)%
      \includegraphics{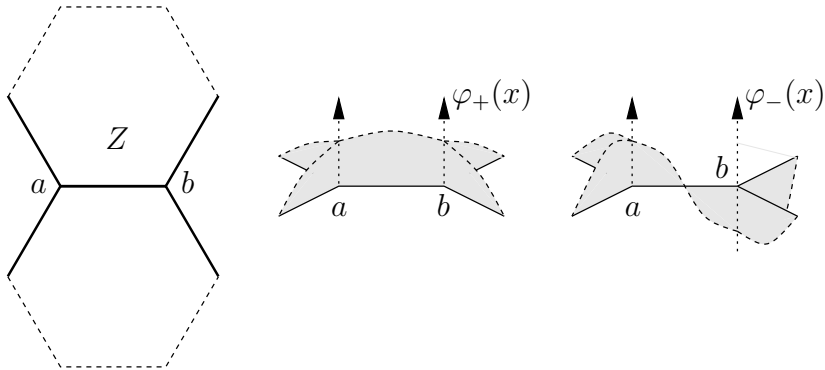}%
    \end{picture}%
    \setlength{\unitlength}{4144sp}%
    \begin{picture}(4759,2179)(159,-1470)
      \put(2836,119){$\phi_+(x)$}%
      \put(4591,119){$\phi_-(x)$}%
      \put(1216,-421){$b$}%
      \put(316,-421){$a$}%
      \put(766,-151){$Z$}%
      \put(2116,-556){$a$}%
      \put(2746,-556){$b$}%
      \put(3871,-556){$a$}%
      \put(4411,-331){$b$}%
    \end{picture}%

    \caption{The extra eigenstates outside the Dirichlet spectrum for
      the zig-zag nano-tubes with even number of ``zig-zags''. On the
      left, the support is shown; on the right, the eigenfunctions
      $\phi_\pm$ corresponding to the smallest solution of
      $\eta(\lambda) = \cos \sqrt \lambda = \pm 1/3$ are plotted (in
      the case of zero potential $q_0=0$).}
    \label{fig:extra-ef}
\end{figure}
It is constructed as follows: on the two ``horns'' directed toward the
vertex $a$, we define the function to be equal to $\phi_{1,\lambda}
(x)$. It is similarly defined on the ``horns'' leading towards $b$. On
the ``bridge'' between $a$ and $b$, we define the function as
$\phi_{0,\lambda}+\phi_{1,\lambda}$. It is easy to conclude that the
equality $\phi^\prime_{1,\lambda}(0)=3\phi^\prime_{1,\lambda}(1)$
implies that the function satisfies Neumann conditions at both
vertices (and certainly the equation $H\phi=\lambda\phi$ on the edges
of $Z$). The graph of the function $\phi_+$ is visualized in the
middle of \Fig{extra-ef}.

Analogously, if $\eta(\lambda)=-1/3$, one creates a function
$\phi_{-}$, changing the value on the bridge to
$\phi_{0,\lambda}-\phi_{1,\lambda}$ (see the right graph in
\Fig{extra-ef}).

The functions $\phi_{\pm}$ are defined on $Z$ only, but can be
extended to the whole structure $G$ as follows: one repeats the
functions up and down (to dashed-line hexagons in \Fig{extra-ef}),
alternating the sign. Outside this column of hexagons, we define the
functions to be equal to zero. These functions are periodic with
respect to the vector $\vec{e}_2$ with period $2$, and thus define
compactly supported eigenfunctions for any even zig-zag nano-tube
$T_{(0,2N)}$. We will call such eigenfunctions the \emph{three-leaf
  eigenfunctions} (the name suggested by the right graph in
\Fig{extra-ef}).

We are now ready to establish the main result about the spectra of
carbon nano-tubes. First of all, let us collect all the notions we
need here. As before, $p=(p_1,p_2)\in\ZZ^2$ is a translation
vector that determines the nano-tube $T_p$. The Hamiltonian $H_p$
on $\Lsqr{T_p}$ is defined as before, using the pull-back of a
potential $q_0(x)$ on $[0,1]$ symmetric with respect to the point
$1/2$. We also denote by $D(\lambda)$ the Hill discriminant (trace
of the monodromy matrix) of the Hill operator $H^\per$ on $\RR$
with periodized potential $q_0$. The subset $B_p$ of the Brillouin
zone $B$ is defined in~\eqref{eq:briou}. Finally, the function
$F(\theta)$ of the quasimomentum $\theta$ is defined
in~\eqref{eq:def.ew.func}, and $\alpha(p)$ is described in
\Lem{restrict}.

In order to avoid lengthy formulation, in the Theorem and two
Corollaries below, when dispersion relations are described, the
flat branches corresponding to the pure point spectrum are
omitted. The pure point spectra (and thus the flat branches) are
described in separate statements.

\begin{theorem}
  \label{thm:spec.nano-tubes}
\indent
\begin{enumerate}
\item The non-constant part of the dispersion relation for $H_p$ is
  provided by
    \begin{equation}
      \label{eq:tube_dispersion}
      D(\lambda)=\pm \frac 23 F(\theta), \qquad \theta \in B_p.
    \end{equation}

  \item The singular continuous spectrum $\spec[sc]{H_p}$ is empty.

   \item The absolutely continuous spectrum is given by
     \begin{equation}
       \label{eq:tube_ac}
       \spec[ac]{H_p} =
           D^{-1}\Bigl(\bigl[-2,-\frac 23\alpha(p)\bigr] \cup
                       \bigl[ \frac 23 \alpha(p), 2 \bigr]\Bigr)
     \end{equation}
     and
     \begin{equation}
       \label{eq:tube_ac2}
       D^{-1} \Bigl(\bigl[-2,-\frac 23\bigr] \cup
                    \bigl[\frac 23,2 \bigr] \Bigr)
             \subseteq \spec[ac]{H_p} \subseteq
             \spec {H^\per}=
             D^{-1}\bigl([-2,2]\bigr).
     \end{equation}

   \item $\spec[ac]{H_p} = \spec{H^\per}$ if and only if $p_1-p_2$ is
     divisible by $3$.

   \item $\spec[ac]{H_p} = D^{-1}\bigl([-2,-\frac 23] \cup [\frac
     23,2]\bigr)$ if and only if $T_p$ is either a $(0,1)$-, or a
     $(0,2)$- zig-zag nano-tube (or equivalent, e.g.~$T_{(1,-1)}$).

   \item Unless $T_p$ is a zig-zag nano-tube with an even number of
     zig-zags (i.e., $T_{(0,2N)}$), one has
     \begin{equation}
     \label{eq:tube_pp}
       \spec[pp]{H_p} =\Sigma^\Dir.
     \end{equation}
     This spectrum consists of one edge of each spectral gap (including
     the closed ones) of $\spec{H^\per}$.  All these eigenvalues are of
     infinite multiplicity and the corresponding eigenspaces are
     spanned by simple loop eigenfunctions (supported on a single
     hexagon) and tube loop eigenfunctions (supported on a loop around
     the tube).

   \item If $T_p$ is a zig-zag nano-tube with an even number of
     zig-zags (i.e., $T_{(0,2N)}$), one has
     \begin{equation}
       \label{eq:tube_pp2}
       \spec[pp]{H_p} = \Sigma^\Dir \cup \Xi,
     \end{equation}
     where
     \begin{equation}
       \label{eq:tube_ppMu}
       \Xi = D^{-1} \Bigl( \pm \frac 23 \Bigr).
     \end{equation}
     The eigenvalues from $\Xi$ are of infinite multiplicity, are
     embedded into $\spec[ac]{H_p}$, and are located two per each band
     of $\spec{H^\per}$. The corresponding eigenspaces are generated
     by the compactly supported three-leaf functions.

     The eigenvalues from $\Sigma^\Dir$ are of infinite multiplicity
     and the corresponding eigenspaces are spanned by simple loop
     eigenfunctions and tube loop eigenfunctions.

   \item If $p_1-p_2$ is divisible by $3$, the ac-spectrum of $H_p$
     has exactly the same gaps as $\spec{H^\per}$. Otherwise, there is
     an additional gap $D^{-1}\bigl((-\frac 23\alpha(p),\frac
     23\alpha(p))\bigr)$ inside each band of the spectrum of $H^\per$.
  
  \end{enumerate}
\end{theorem}
\begin{proof}
  \indent

  The first statement coincides with~\eqref{eq:dispersion_p}.

  The claim (ii) is proven exactly as the corresponding statement in
  \Thm{spec.graphen}.

  Statements~(iii) through~(v) follow from~(i) and \Lem{restrict}.

  The statement~(vi) is almost completely proven, except the
  description of the eigenfunctions. The proof of this description
  works exactly like in \Thm{spec.graphen}, except that the procedure
  of eliminating hexagons does not have to end with an empty set. One
  can end up with a loop of edges around the tube, which thus does not
  encircle any hexagons. This would provide a loop eigenfunction that
  runs around the tube, rather than around a hexagon. The similar
  claim in~(vii) concerning the eigenvalues from $\Sigma^\Dir$ is
  proven exactly the same way.

  What remains to be proven in~(vii), is the structure of the
  eigenfunctions corresponding to $\lambda\in\Xi$. It is proven
  similarly to the structure of eigenfunctions corresponding to
  $\Sigma^\Dir$. Indeed, again according to~\cite{kuchment:05}, the
  eigenspace is spanned by compactly supported eigenfunctions.
  Consider the outer boundary of the support and start eliminating
  hexagons inside as follows. There must be a vertex (like the ends of
  horns in \Fig{extra-ef}) that borders zero values. Then, on the
  corresponding horn the function must be proportional to
  $\phi_{1,\lambda}$. Let us now extend it to a three-leaf
  eigenfunction and subtract from the original one. Continuing this
  process, we eventually eliminate all hexagons. Notice that in this
  case we cannot end up with a loop around the tube, since this would
  force all the vertex values to be equal to zero, which is
  impossible, when $\lambda$ does not belong to $\Sigma^\Dir$. Thus,
  only the three-leaf states enter the eigenfunction.

  The statement (viii) follows from the previous ones.
\end{proof}

We will specify this result for the cases of zig-zag ($p=(0,N)$) and
armchair ($p=(N,N)$) nano-tubes. The zig-zag case was also considered
in \cite{korotyaev:06}.
\begin{corollary}
  \label{cor:zigzag}
  \indent
  Let $T_{(0,N)}$ be a zig-zag nano-tube with $N$ zig-zags.
  \begin{enumerate}
  \item The non-constant part of the dispersion relation for
    $H_{(0,N)}$ is given by
    \begin{equation}
      \label{eq:zigzag.disp}
      D(\lambda) = \pm \frac 23
     \sqrt {1+4\cos \frac {\pi n} N
        \Bigl[(\cos \frac{\pi n} N  +
               \cos \Bigl(\theta_1 - \frac{\pi n} N \Bigr)
        \Bigr]},
    \end{equation}
    where $0 \le n < N$.
  \item The singular continuous spectrum is empty.
  \item The absolutely continuous spectrum is given by
     \begin{equation}
       \label{eq:zigzag_ac2}
       \spec[ac]{H_{(0,N)}} =
       D^{-1} \Bigl(\bigl[-2, -\frac 23 \alpha \bigr]
         \cup
                    \bigl[\frac 23 \alpha, 2 \bigr] \Bigr)
     \end{equation}
     where $\alpha=\alpha((0,N)) \in [0,1]$ is defined
     in~\eqref{eq:alpha_0N}. In particular, $\alpha=0$ (i.e.,
     $\spec[ac]{H_{(0,N)}} = \spec{H^\per}$) if and only if $N$ is
     divisible by $3$.  Furthermore, $\alpha=1$ if and only if $N=1$
     or $N=2$.
   \item \sloppy If $N$ is odd, then the pure point spectrum is given
     by $\spec[pp]{H_{(0,N)}} = \Sigma^\Dir$. The corresponding eigenspaces
     are infinite-dimensional and generated by simple loop
     eigenfunctions (supported on a single hexagon) and tube loop
     eigenfunctions (supported on a loop around the tube).

     If $N$ is even, then $\spec[pp]{H_{(0,N)}} = \Sigma^\Dir \cup
     \Xi$ where $\Xi$ is defined in~\eqref{eq:tube_ppMu}. In
     particular, if $N=2$ then the embedded eigenvalues from $\Xi$ are
     located at the band edges of $\spec[ac]{H_p}$. If $N>2$ is even,
     this new point spectrum is located inside the bands. The
     eigenspaces corresponding to $\Xi$ are infinite-dimensional and
     generated by the compactly supported three-leaf functions.

   \item $\spec{H_{(0,N)}}$ has additional gaps (other than the gaps
     of $\spec{H^\per}$) if and only if $N$ is not divisible by $3$.
  \end{enumerate}
\end{corollary}

\begin{remark}
  In order to avoid confusion, we need to specify what a simple loop
  eigenstate is for the case of the necklace tube $T_{(0,1)}$. In this
  case, the image of a hexagon in the tube is a ``dumbbell''
  consisting of two beads of the necklace connected with a link (see
  \Fig{link} below).
\begin{figure}[ht]
    \centering \includegraphics[scale=0.5]{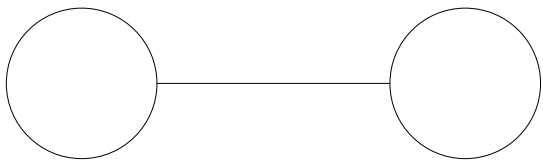}
    \caption{A dumbbell image of a hexagon.}
      \label{fig:link}
  \end{figure}
  A simple loop eigenfunction in this case can concentrate either on a
  single bead, or on the whole dumbbell.
\end{remark}

\begin{corollary}
  \label{cor:armchair}
  \indent
  Let $T_{(N,N)}$ be an armchair nano-tube.
  \begin{enumerate}
  \item The non-constant part of the dispersion relation for
    $H_{(N,N)}$ is given by
    \begin{equation}
      \label{eq:armchair.disp}
      D(\lambda) = \pm \frac 23
     \sqrt {1+8\cos \Bigl(\theta_1 - \frac {\pi n} N \Bigr)
               \cos \Bigl(\frac{\theta_1} 2 \Bigr)
               \cos \Bigl(\frac {\theta_1}2 - \frac {\pi n} N \Bigr)},
    \end{equation}
    where $0 \le n < N$.
  \item The singular continuous spectrum is empty.
  \item The absolutely continuous spectrum is given by
     \begin{equation}
       \label{eq:armchair_ac2}
       \spec[ac]{H_{(N,N)}} =
       D^{-1} \bigl([-2, 2 ] \bigr) = \spec{H^\per}.
     \end{equation}
   \item The pure point spectrum is given by $\spec[pp]{H_{(N,N)}} =
     \Sigma^\Dir$ and is located at an edge of each gap in
     $\spec[ac]{H_{(N,N)}}$. The eigenvalues are of infinite
     multiplicity and the eigenspaces are generated by simple loop
     eigenfunctions (either on a single hexagon or a loop around the
     tube).
   \item $\spec{H_{(N,N)}}$ has exactly the same gaps as
     $\spec{H^\per}$.
  \end{enumerate}
\end{corollary}

\section{Final remarks}
\label{sec:outlook}
\begin{itemize}

\item The zig-zag nano-tube case has been thoroughly studied by a
  different method in \cite{korotyaev:06, korotyaev:06_magn}. The
  methods employed in this paper seem to be significantly simpler than
  the ones in \cite{korotyaev:06, korotyaev:06_magn} and also apply to
  all $2D$ carbon nano-structures such as graphene and any single-wall
  nano-tube.

  After the paper was accepted for publication, the authors received
  the preprint \cite{Kor_arm}, where the case of armchair nano-tubes is
  considered by methods analogous to the ones of \cite{korotyaev:06,
    korotyaev:06_magn}.

\item Other spectral properties of the operators $H$ and $H_p$, e.g.
  asymptotics of gaps lengths or properties of (and formulas for) the
  density of states can be easily derived from the explicit dispersion
  relations that we obtained and the well studied properties of the
  Hill discriminant. As an example, we provide a theorem below that
  describes the smoothness of the potential in terms of the gap decay.
  In order to do so, we call the gaps arising as $D^{-1}([-\frac
  23\alpha (p), \frac 23\alpha(p)])$ the \textit{odd gaps} $G_{2k-1},
  k=1, 2, ...$ and the gaps of the Hill operator the \textit{even
    gaps} $G_{2k}, k=1, 2, ...$.  Notice that we count the gaps even
  when they close (e.g., all odd gaps close for graphene and for
  nano-tubes with integer $(p_1-p_2)/3$). Let also $\gamma_k$ be the
  lengths of the gap $G_k$.

  In the theorem below, the operator is either the graphene operator
  $H$, or the nano-tube operator $H_p$.
  \begin{theorem}
    \begin{enumerate} 
    \item The periodized $1D$ potential $q_0$ is infinitely
      differentiable if and only if $\gamma_{2k}$ decays faster than
      any power of $k$ when $k\rightarrow \infty$.
    \item The periodized $1D$ potential $q_0$ is analytic if and only
      if $\gamma_{2k}$ decays exponentially with $k$.
    \end{enumerate} 
  \end{theorem}
  Since the even gaps are exactly the spectral gaps for the Hill
  operator with the periodized $q_0$, this is an immediate corollary
  of the results of this text and known results of the same nature for
  the Hill operator \cite{Mityagin, Hoch1, Hoch2, Laz, Trub}.
  Statements similar to this theorem can be derived as easily for
  other functional classes of potentials, using the corresponding
  results for the Hill operator in \cite{Mityagin}.

\item Albeit for graphene (as well as for the nano-tubes with
  $p_1-p_2$ divisible by $3$) the absolutely continuous spectrum
  coincides with the one of the periodic Hill operator as a set, the
  structure of the spectrum is different, due to the appearence of the
  conical singularities inside of each band of the Hill spectrum.

  Such singularities can also appear when the even gaps close, but
  this situation is non-generic with respect to the potential $q_0$.
  However, as we discussed above, closing the odd gaps and thus
  appearence of conical singularities there is mandatory for any
  potential in the graphene case, as well for nano-tubes $T_p$ with
  $p_1-p_2$ divisible by $3$.

\item As we have indicated in the beginning, quantum graphs (quantum
  networks) have been used to model spectra of molecules at least
  since \cite{Pau, RuS}. However, the validity of such models is still
  under investigation, see e.g., \cite{DuclosExner, ExSeba_challenge,
    Kuch_thin, MolVai, MolVai2, Post_decoupled} and references
  therein.

\item The graphene operator $H$ provides also an interesting example in
  terms of Liouville type theorems. As it was established in
  \cite{KuPinch2} (see also \cite{KuPinch1} for related
  considerations), the Liouville theorem for periodic operators holds
  if and only if the Fermi surface consists of finitely many points.
  Albeit this normally occurs at the spectral edges only, it was
  indicated in \cite{KuPinch1,KuPinch2} that in principle this can
  happen inside the spectrum.  The graphene operator provides just
  such an example. Namely, a direct corollary of our results and the
  ones of \cite{KuPinch2} is the following
  \begin{theorem} Let $H$ be the graphene operator. Suppose $D(\lambda)=0$ and 
$n>0$. Then the space of solutions $u$ of the
    equation
    \begin{equation*}
      Hu-\lambda u=0
    \end{equation*}
    such that
    \begin{equation*}
      |u(x)|\leq C_u (1+|x|)^n
    \end{equation*}
    is finite dimensional.  
  \end{theorem}
\end{itemize}
\section*{Acknowledgments}
The authors express their gratitude to E.~Korotyaev, K.~Pankrashkin
and V.~Pokrovsky for information and comments. In particular, it was
V.~Pokrovsky, who attracted our attention to the importance of
conical singularities.

The authors are also grateful to the reviewer for useful remarks.

This research of both authors was partly sponsored by the NSF through
the NSF Grant DMS-0406022. The authors thank the NSF for this support.
The second author was partly supported by the DFG through the
Grant~Po~1034/1-1. Part of this work was done during O.~P.\ visiting
Texas A\&M University. The second author thanks the DFG for this
support and Texas A\&M University for the hospitality.


\end{document}